\documentstyle[aps]{revtex}

\begin{document}

\title{ON SOME PECULIARITIES OF SOLVING NONSTATIONARY
PROBLEMS OF QUANTUM MECHANICS}
\author{V.\,E.\, Mitroshin\footnote{
mitroshin@univer.kharkov.ua}}
\address{Kharkov National University,Ukraine,61077,Kharkov}

\date{April 27, 2001; revised version June 25, 2001}
\maketitle

\vspace{0.5cm}
\begin{abstract}
Exact solutions of several nonstationary problems of quantum
mechanics are obtained. It is shown that if the initial
conditions of the problem correspond to the localized-in-space
particle, then it moves exactly along the classical trajectory,
and the wave packet is not spread in time.
\end{abstract}
\vspace{0.3cm}
\hspace{1.35cm}PACS number(s): 03, 03.65.Ca, 03.65.Ge, 03.65.Nk

\section{Introduction}
The standard way of solving a nonstationary problem of quantum
mechanics (in $\hbar = m = e = 1$ units)
\begin{eqnarray*}
i\frac{\partial }{{\partial t}}\varphi \left( {t,x} \right) =
\left( {H_0 \left( x \right) + H_{{\mathop{\rm int}} } \left( {t,x}
\right)} \right)\varphi \left( {t,x} \right),
\end{eqnarray*}
with the initial condition $
\varphi (t,x)|_{t = 0}  = k(x)
$, where $H_o$ is the nonperturbed Hamiltonian, and $H_{\rm
int}$ is the time-dependent perturbation, lies in the
transition to the representation of the interaction $\varphi
\left( {t,x}
\right) = \exp \left( { - itH_0 } \right)\psi \left( {t,x}
\right)$, and for the new equation
\begin{equation}\label{eq1}
i\frac{\partial }{{\partial t}}\psi \left( {t,x} \right) = \exp
\left( {itH_0 } \right)H_{{\mathop{\rm int}} } \left( {t,x}
\right)\exp \left( { - itH_0 } \right)\psi \left( {t,x} \right)
\end{equation}
powerful methods of perturbation theory are developed, the
presentation of which can be found nearly in any textbook on
quantum mechanics. But here we would like to draw the readers'
attention to a number of exactly solvable problems for
eq.(\ref{eq1}). It has been found that if an infinitely
differentiated function with a compact carrier is chosen as an
initial condition, then the wave packet not only moves along
the trajectory exactly corresponding to the solution of similar
problems of classical mechanics, but is also not spread in
time.
\section{Procedure}
We make use of the fact that
\begin{equation}\label{eq2}
e^{itH_0 } H_{{\mathop{\rm int}} } e^{ - itH_0 } =
H_{{\mathop{\rm int}} } + \sum\limits_{n = 1}
{\frac{1}{{n!}}\left( {it} \right){}^n} ad^n H_0 \left(
{H_{{\mathop{\rm int}} } } \right),
\end{equation}
where $ad^n H_0 (H_{{\mathop{\rm int}} } ) = [H_0 [H_0
...[H_0 ,H_{{\mathop{\rm int}} } ]..]
$ is a contracted form of $n$-fold commutator. It remains to
choose such $H_{\rm int}$, at which the series (\ref{eq2}) is
either summable or stops at a certain term. For simplicity, we
shall consider one-dimensional problems in order not to make
calculations cumbersome. In the subsequent discussion we shall
often refer to two special functions with proper names being
reserved for them: the Heaviside function and the Kelly
function. The Heaviside function is the function $\chi \left(
t\right) = 0$ at $t<0$ and $
\chi \left( t \right) = 1
$ at $t\geq0$. The Kelly function is the function
\[ {\cal K}_{\alpha
,\beta } \left( x \right) = const \times \exp \left[ {{{-1}
\mathord{\left/
 {\vphantom {{ - 1} {\left( {x - \alpha } \right)\left( {\beta  - x} \right)}}} \right.
 \kern-\nulldelimiterspace} {\left( {x - \alpha } \right)\left( {\beta  - x} \right)}}} \right]
\]
at $\alpha < x < \beta $ and equal to zero for all other
argument values. This function is remarkable for its being
nonzero only within a small interval
$\left({\alpha,\beta}\right)$ and having derivatives of any
order, square integrable by the Lebesgue method.

So far as operator $u(t,x)=\exp(itH_o)$ is a unitary operator
we obtain
 that the function $\varphi=\exp(itH_o)\psi\left(t,x\right)$ is being nonzero only within a
interval where the function $\psi\left(t,x\right)$ is nonzero.
And now we turn to particular problems.

\section{A freely travelling particle}
It would appear surprising to find anything new in this problem
investigated so thoroughly. And yet, has the problem been posed
correctly if we wish to describe the motion of the particle
which acquires a velocity $\lambda$ at the time moment $t=0$?
In other words, we should solve, in fact, equation (\ref{eq1})
with the
perturbation proportional to the momentum operator, i.e., of
the form $H_{{\mathop{\rm int}} } = - i\lambda \theta
(t){\partial \mathord{\left/
 {\vphantom {\partial  {\partial x}}} \right.
 \kern-\nulldelimiterspace} {\partial x}}
$, where $\theta (t) $ is a certain function of time defining
how exactly the momentum was transferred to the particle, and
$\lambda$ is the constant characterizing the momentum transfer
value, its dimensional representation is such that should have
the dimensionality inverse to time. Since $H_0 = - \left( {{1
\mathord{\left/
 {\vphantom {1 2}} \right.
 \kern-\nulldelimiterspace} 2}} \right){{\partial ^2 } \mathord{\left/
 {\vphantom {{\partial ^2 } {\partial x^2 }}} \right.
 \kern-\nulldelimiterspace} {\partial x^2 }}
$ commutes with $H_{\rm int}$, then eq.(\ref{eq1}) takes on the
following form
\begin{equation}\label{eq3}
i\frac{\partial }{{\partial t}}\psi \left( {t,x} \right) = -
i\lambda \theta (t)\frac{\partial }{{\partial x}}\psi \left(
{t,x} \right).
\end{equation}
The general solution to this equation is \[
\psi (t,x) = k(x - \lambda \int\limits_0^t {\theta \left( \tau  \right)d\tau } )
\]
for $t\geq0$, and $k$ is any infinitely differentiated function
of argument satisfying the initial requirements. In particular,
any infinitely differentiated function with a compact carrier,
equal identically to zero beyond a certain interval, e.g., the
Kelly function, can be taken as $k$. If the particle instantly
acquires the velocity $\lambda$, the one should put $
\theta (t) = \chi \left( t \right)
$. As a result, we obtain that the probability density (in the
representation of the interaction) for the particle to be at a
certain point of space-time $|\psi (x,t)|^2 = {\cal K}_{\alpha
,\beta }^2 (x - \lambda t)$ is not spread in time and it
proceeds exactly along the trajectory described by classical
mechanics. If, however, the particle gained its velocity at
uniform acceleration, i.e., $\lambda
\theta (t) = a
\times t $, then $
|\psi(x,t)|^2 = {\cal K}_{\alpha ,\beta }^2 (x - a \times t^2
/2) $, this again exactly corresponds to the solution of
classical mechanics, as well as $|\varphi(x,t)|^2=
|u(t,x)\psi(x,t)|^2 $.

\section{A charged particle in the electric field}
$H_0 = - \left( {{1 \mathord{\left/
 {\vphantom {1 2}} \right.
 \kern-\nulldelimiterspace} 2}} \right){{\partial ^2 } \mathord{\left/
 {\vphantom {{\partial ^2 } {\partial x^2 }}} \right.
 \kern-\nulldelimiterspace} {\partial x^2 }},
$ and the electric field $H_{{\mathop{\rm int}} } = \lambda
\theta \left( t \right)x
$ included by the $\theta (t)$--law. Since $[H_0
,H_{{\mathop{\rm int}} } ] = - \lambda \theta (t){\partial
\mathord{\left/
 {\vphantom {\partial  {\partial x}}} \right.
 \kern-\nulldelimiterspace} {\partial x}}
$, then $[H_0 [H_0 ,H_{{\mathop{\rm int}} } ]] = 0 $, and
eq.(\ref{eq1}) takes the form:
\[
i\frac{\partial }{{\partial t}}\psi \left( {t,x} \right) =
\lambda \theta \left( t \right)x\psi (t,x) - i\lambda t\theta (t)\frac{\partial }{{\partial x}}\psi (t,x)
\]
After the replacement $\psi (t,x) = \rho (t,x)\exp \{ -
i\lambda x\int\limits_0^t {\theta (\tau )d\tau + i\lambda ^2
\int\limits_0^t {\tau \theta (\tau )} [} \int\limits_0^\tau
{\theta (\tau ')d\tau ']d\tau \} } $, we obtain the following
equation
\begin{equation}\label{eq4}
\frac{\partial }{{\partial t}}\rho (t,x) =  - \lambda t\theta (t)\frac{\partial }{{\partial x}}\rho
(t,x),
\end{equation}
the general solution to which is the function $
\rho (t,x)=k(x-\lambda \int\limits_0^t {\tau \theta (\tau )d\tau })
$. So, we have that if the Heaviside function (the electrical
field is included instantly) is chosen as $\theta(t)$, and the
Kelly function is chosen at $t=0$ as an initial condition, then
the probability density precisely follows the classical path
$|\psi (x,t)|^2 = {\cal K}_{\alpha ,\beta }^2 (x - \lambda
\times t^2 /2)
$, and the wave packet is not spread in time as well as
$|\varphi(x,t)|^2= |u(t,x)\psi(x,t)|^2 $.

\section{A harmonic oscillator in the electric field}
In this case, we put $H_{{\mathop{\rm int}} } = \lambda
\theta \left( t \right)x$ and $H_0 =-\left({{1 \mathord{\left/
 {\vphantom {1 2}} \right.
 \kern-\nulldelimiterspace} 2}} \right){{\partial ^2 } \mathord{\left/
 {\vphantom {{\partial ^2 } {\partial x^2 }}} \right.
 \kern-\nulldelimiterspace} {\partial x^2 }} + 1/2x^2
$. Let us calculate the commutators.

As $[H_0 ,H_{{\mathop{\rm int}} } ] = - \lambda \theta
(t){\partial \mathord{\left/
 {\vphantom {\partial  {\partial x}}} \right.
 \kern-\nulldelimiterspace} {\partial x}}
$, then $[H_0 [H_0 ,H_{{\mathop{\rm int}} } ]] = \lambda
\theta (t)x = H_{{\mathop{\rm int}} } $, and the series (\ref{eq2})
is summed to $
\lambda x\theta (t)\cos t - i\lambda \theta (t)\sin t{\partial  \mathord{\left/
 {\vphantom {\partial  {\partial x}}} \right.
 \kern-\nulldelimiterspace} {\partial x}}
$. Then, eq.(\ref{eq1}) is written as
\begin{equation}\label{eq5}
i\frac{\partial }{{\partial t}}\psi (t,x) =
\lambda x\theta (t)\cos t\psi (t,x) - i\lambda \theta (t)\sin t\frac{\partial }{{\partial x}}\psi (t,x)
\end{equation}
and can be integrated
explicitly:
\[
\psi (t,x) = \kappa (x - \lambda \int\limits_0^t {\theta (\tau )\sin \tau d\tau } )
\times \exp \{  - i\lambda x\int\limits_0^t {\theta (\tau )\cos \tau d\tau }  + i\lambda ^2 \int\limits_0^t {\theta (\tau )\sin \tau [\int\limits_0^\tau  {\theta (\tau ')\cos \tau 'd\tau
']d\tau\}}}. \]If the electrical field is included instantly,
$
\theta (t) = \chi \left( t \right)
$, then for the probability density we obtain the answer
exactly corresponding to the solution of a similar problem in
classical mechanics: $|\psi (x,t)|^2 = {\cal K}_{\alpha ,\beta
}^2 (x + 2\lambda \sin ^2 t/2)$ as well as $|\varphi(x,t)|^2=
|u(t,x)\psi(x,t)|^2 $.

One can consider the pendulum
swinging under the action of force, or the pendulum acquiring
the momentum at a certain moment of time, etc., all these
problems have the solutions differing from the "classical
solution" only by the presence of a complex phase factor, whose
value is very much similar to the integral over the pendulum
trajectory ("intrinsic" clock that is "ticking" as long as the
perturbation acts).

\section{Conclusion}
The given problems are far from being the only ones that have
exact analytical solutions, the solutions correlating well with
the solutions to similar problems of classical mechanics. And
in view of this many questions arise. What is "the freely
travelling particle"? In fact, once it had been given the
momentum to "travel"! And what is "the particle in the electric
field"? The field was "turned on" at one time, after all! And
the equation $i\dot \varphi _t = -
{\raise0.5ex\hbox{$\scriptstyle 1$}
\kern-0.1em/\kern-0.15em
\lower0.25ex\hbox{$\scriptstyle 2$}}\ddot \varphi _{xx}
$ is only an equation of free motion, it has no physical
meaning in itself. The Fourier transform of this equation shows
only how the equation looks like in the vector space $
\{ e{}^{ - ikx}\}
$, and nothing more. Being pure materialists, we impart a
desirable for us "physical meaning" to them, similarly to
Bernoulli who, at his time, treated the possibility of
Fourier-series expansion of the function due to "world
harmony". We would like to discuss the philosophy of quantum
mechanics in greater detail - thousands of books have been
written on the topic, so many passions were burning! But the
time has not come as yet, not all the problems (passage through
a barrier, interference) are solved, that might be helpful to
open the mysterious meaning of the "wave-particle" dualism, the
more so - it is not the right place.

\end{document}